\documentclass{article}

\usepackage{graphicx} 
\usepackage{subfigure} 

\usepackage{natbib}

\usepackage{algorithm}
\usepackage{algorithmic}

\usepackage{hyperref}

\usepackage[accepted]{icml2013}

\usepackage{float}

\icmltitlerunning{Extractive Summarization of Electronic Health Record Discharge Notes}

\begin{document} 

\twocolumn[
\icmltitle{Extractive Summarization of EHR Discharge Notes}

\icmlauthor{Emily Alsentzer}{emilya@mit.edu}

\icmlauthor{Anne Kim}{anneykim@mit.edu}


\vskip 0.3in
]

\begin{abstract} 
Patient summarization is essential for clinicians to provide coordinated care and practice effective communication. Automated summarization has the potential to save time, standardize notes, aid clinical decision making, and reduce medical errors. Here we provide an upper bound on extractive summarization of discharge notes and develop an LSTM model to sequentially label topics of history of present illness notes. We achieve an F1 score of 0.876, which indicates that this model can be employed to create a dataset for evaluation of extractive summarization methods.
\end{abstract} 

\section{Introduction}
Summarization of patient information is essential to the practice of medicine. Clinicians must synthesize information from diverse data sources to communicate with colleagues and provide coordinated care. Examples of clinical summarization are abundant in practice; patient handoff summaries facilitate provider shift change, progress notes provide a daily status update for a patient, oral case presentations enable transfer of information from overnight admission to the care team and attending, and discharge summaries provide information about a patient's hospital visit to their primary care physician and other outpatient providers \cite{1Feblowitz2011}.

Informal, unstructured, or poor quality summaries can lead to communication failures and even medical errors, yet clinical instruction on how to formulate clinical summaries is ad hoc and informal. Non-existent or limited search functionality, fragmented data sources, and limited visualizations in electronic health records (EHRs) make summarization challenging for providers \cite{2Christensen208, 3Singh2013, 4Natarajan2010}. Furthermore, while dictation of EHR notes allows clinicians to more efficiently document information at the point of care, the stream of consciousness-like writing can hinder the readability of notes. Kripalani et al. show that discharge summaries are often lacking key information, including treatment progression and follow-up protocols, which can hinder communication between hospital and community based clinicians \cite{5Kripalani2007}. Recently, St. Thomas Hospital in Nashville, TN stipulated that discharge notes be written within 48 hours of discharge following incidences where improper care was given to readmitted patients because the discharge summary for the previous admission was not completed\cite{alsentzer}. 

Automated summary generation has the potential to save clinician time, avoid medical errors, and aid clinical decision making. By organizing and synthesizing a patient's salient medical history, algorithms for patient summarization can enable better communication and care, particularly for chronically ill patients, whose medical records often contain hundreds of notes. In this work, we explore the automatic summarization of discharge summary notes, which are critical to ensuring continuity of care after hospitalization. We (1) provide an upper bound on extractive summarization by assessing how much information in the discharge note can be found in the rest of the patient's EHR notes and (2) develop a classifier for labeling the topics of history of present illness notes, a narrative section in the discharge summary that describes the patient's prior history and current symptoms. Such a classifier can be used to create topic specific evaluation sets for methods that perform extractive summarization. These aims are critical steps in ultimately developing methods that can automate discharge summary creation. 

\section{Related Work}
In the broader field of summarization, automization was meant to standardize output while also saving time and effort. Pioneering strategies in summarization started by extracting "significant" sentences in the whole corpus to build an abstract where "significant" sentences were defined by the number of frequently occurring words \cite{Luhn1958}. These initial methods did not consider word meaning or syntax at either the sentence or paragraph level, which made them crude at best. More advanced extractive heuristics like topic modeling \cite{allahyari2016semantic}, cue word dictionary approaches \cite{Edmundson1969cue}, and title methods \cite{ferreira2013titlemethod} for scoring content in a sentence followed soon after. For example, topic modeling extends initial frequency methods by assigning topics scores by frequency of topic signatures, clustering sentences with similar topics, and finally extracting the centroid sentence, which is considered the most representative sentence \cite{allahyari2017}.  Recently,  abstractive summarization approaches using sequence-to-sequence methods have been developed to generate new text that synthesizes original text\cite{paulus2017deep, nallapati2016abstractive, tensorflow, rush2015neural}; however, the field of abstractive summarization is quite young.

Existing approaches within the field of electronic health record summarization have largely been extractive and indicative, meaning that summaries point to important pieces in the original text rather than replacing the original text altogether. Few approaches have been deployed in practice, and even fewer have demonstrated impact on quality of care and outcomes \cite{8Pivovarov2016}. Summarization strategies have ranged from extraction of “relevant” sentences from the original text to form the summary \cite{7Moen2016}, topic modeling of EHR notes using Latent Dirichlet allocation (LDA) or bayesian networks \cite{8Pivovarov2016}, and knowledge based heuristic systems \cite{9Goldstein2017}. To our knowledge, there is no literature to date on extractive or abstractive EHR summarization using neural networks. 




\section{Methods}

\subsection{Data}
MIMIC-III is a freely available, deidentified database containing electronic health records of patients admitted to an Intensive Care Unit (ICU) at Beth Israel Deaconess Medical Center between 2001 and 2012. The database contains all of the notes associated with each patient's time spent in the ICU as well as 55,177 discharge reports and 4,475 discharge addendums for 41,127 distinct patients. Only the original discharge reports were included in our analyses. Each discharge summary was divided into sections (Date of Birth, Sex, Chief Complaint, Major Surgical or Invasive Procedure, History of Present Illness, etc.) using a regular expression. 
\subsection{Upper Bound on Summarization}
Extractive summarization of discharge summaries relies on the assumption that the information in the discharge summary is documented elsewhere in the rest of the patient's notes. However, sometimes clinicians will document information in the discharge summary that may have been discussed throughout the hospital visit, but was never documented in the EHR. Thus, our first aim was to determine the upper bound of extractive summarization.

For each patient, we compared the text of the discharge summary to the remaining  notes for the patient's current admission as well as their entire medical record. Concept Unique Identifiers (CUIs) from the Unified Medical Language System (UMLS) were compared in order to assess whether clinically relevant concepts in the discharge summary could be located in the remaining notes \cite{cuis}. CUIs were extracted using Apache cTAKES \cite{savova2010ctakes} and filtered by removing the CUIs that are already subsumed by a longer spanning CUI. For example, CUIs for "head" and "ache" were removed if a CUI existed for "head ache" in order to extract the most clinically relevant CUIs. 

In order to understand which sections of the discharge summaries would be the easiest or most difficult to summarize, we performed the same CUI overlap comparison for the  chief complaint, major surgical or invasive procedure, discharge medication, and history of present illness sections of the discharge note separately. We calculated which fraction of the CUIs in each section were located in the rest of the patient's note for a specific hospital stay. We also calculated what percent of the genders recorded in the discharge summary were also recorded in the structured data for the patient.

\subsection{Labeling History of Present Illness Notes}
\subsubsection{Annotation}
We developed a classifier to label topics in the history of present illness (HPI) notes, including demographics, diagnosis history, and symptoms/signs, among others.
A random sample of 515 history of present illness notes was taken, and each of the notes was manually annotated by one of eight annotators using the software Multi-document Annotation Environment (MAE) \cite{mae}. MAE provides an interactive GUI for annotators and exports the results of each annotation as an XML file with text spans and their associated labels for additional processing. 40\% of the HPI notes were labeled by clinicians and 60\% by non-clinicians. Table \ref{annotation_table} shows the instructions given to the annotators for each of the 10 labels. The entire HPI note was labeled with one of the labels, and instructions were given to label each clause in a sentence with the same label when possible.

\begin{table}
\caption{HPI Categories and Annotation Instructions}
\vspace{15pt}
\label{annotation_table}
\noindent\begin{tabular}{|p{3.1cm}|p{4.5cm}|} \hline
Demographics & Age, gender, race/ethnicity, language, social history etc.  \\
\hline
DiagnosisHistory & Diagnoses prior to current symptoms
\\ \hline
MedicationHistory & Medications or treatment prior to current symptoms \\
\hline
ProcedureHistory & Procedures prior to current symptoms \\ \hline
Symptoms/Signs & Current Signs/Symptoms; chief complaint \\ \hline
Vitals/Labs & Any vitals or labs related to current care episode\\ \hline
Procedures/Results & Procedures including CT, CXR, etc. and their results \\ \hline
Meds/Treatments & Medication or treatment administered in current care episode\\ \hline
Movement & Description of  patient admitted/transferred between hospitals or care units\\ \hline
Other & Does not fit into any other category \\ \hline
\end{tabular}
\end{table}

\subsubsection{Model}
\begin{figure}[h!]
\includegraphics[scale = 0.48]{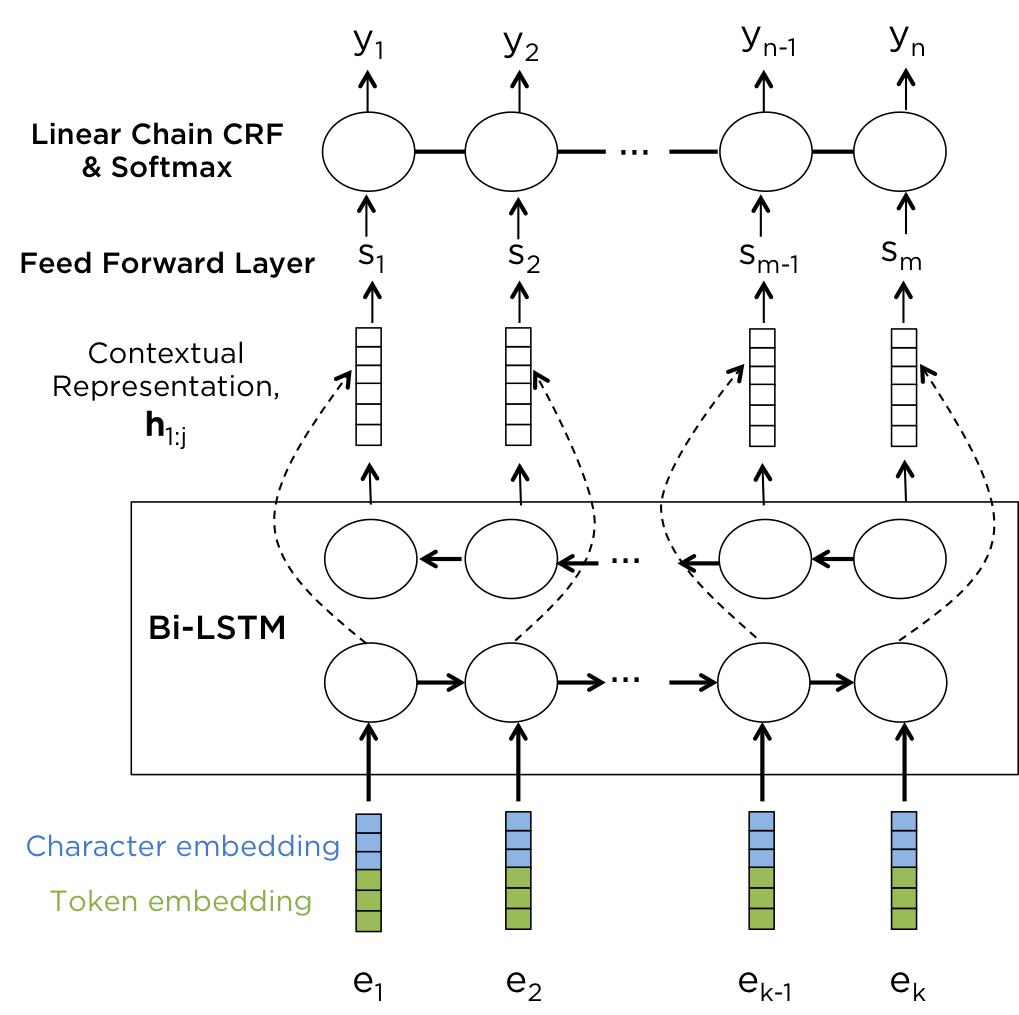}\\
\vspace{10pt}

\includegraphics[scale = 0.48]{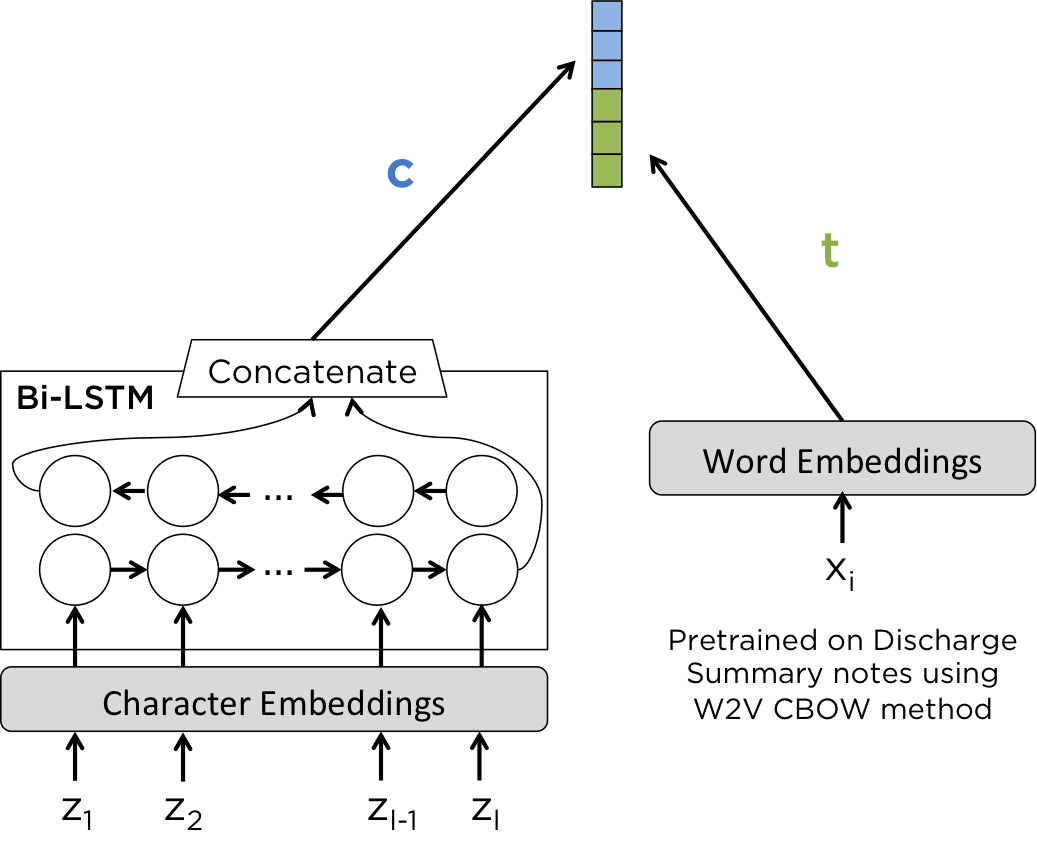}

\centering
\caption{LSTM Model for sequential HPI word classification. $X_i$: token $i$, $Z_1$ to $Z_l$: characters of $X_i$, $\textbf{t}$: word embeddings, $\textbf{c}$: character embeddings, $e_{1:k}$: hybrid token embeddings, $\textbf{h}_{1:j}$: contextual representation embedings, $\textbf{s}_{1:m}$}: score vectors where \textbf{s}[i] is the score for category i, $y_{1:m}$: output word labels.
\label{lstm_model}
\end{figure}

Our LSTM model was adopted from prior work by Dernoncourt et al \cite{10Dernoncourt2016}. Whereas the Dernoncourt model jointly classified each sentence in a medical abstract, here we jointly classify each word in the HPI summary. Our model consists of four layers: a token embedding layer, a word contextual representation layer, a label scoring layer, and a label sequence optimization layer (Figure \ref{lstm_model}).

In the following descriptions, lowercase italics is used to denote scalars, lowercase bold is used to denote vectors, and uppercase italics is used to denote matrices. 

\textbf{Token Embedding Layer:} In the token embedding layer, pretrained word embeddings are combined with learned character embeddings to create a hybrid token embedding for each word in the HPI note. The word embeddings, which are direct mappings from word $x_i$ to vector $\textbf{t}$, were pretrained using word2vec \cite{mikolov2013word2vec, mikolov2013distributed, mikolov2013linguistic} on all of the notes in MIMIC (v30) and only the discharge notes. Both the continuous bag of words (CBOW) and skip gram models were explored. 

Let $z_{1:l}$ be the sequence of characters comprising the word $x_i$. Each character is mapped to its embedding $\textbf{c}_i$, and all embeddings are input into a bidirectional LSTM, which ultimately outputs $\textbf{c}$, the character embedding of the word $x_i$.

The output of the token embedding layer is the vector \textbf{e}, which is the result of concatenation of the word embedding, \textbf{t}, and the character embedding, \textbf{c}.

\textbf{Contextual Representation Layer:} The contextual representation layer takes as input the sequence of word embeddings, $\textbf{e}_{1:k}$, and outputs an embedding of the contextual representation for each word in the HPI note. The word embeddings are fed into a bi-directional LSTM, which outputs $h_j$, a concatenation of the hidden states of the two LSTMs for each word. 

\textbf{Label Scoring Layer:} At this point, each word $x_i$ is associated with a hidden representation of the word, $h_j$. In the label scoring layer, we use a fully connected neural network with one hidden layer to output a score associated with each of the 10 categories for each word. Let $W \in R^{10Xk} $ and $b \in R^{10}$. We can compute a vector of scores \textbf{s} = $W  \cdot h + b$ where the ith component of \textbf{s} is the score of class i for a given word. 

\textbf{Label Sequence Optimization Layer:} The Label Sequence Optimization Layer computes the probability of a labeling sequence and finds the sequence with the highest probability. In order to condition the label for each word on the labels of its neighbors, we employ a linear chain conditional random field (CRF) to define a global score, $Q$, for a sequence of words and their associated scores $s_1,...,s_m$ and labels, $y_1,...,y_n$:
\begin{equation}
\label{eq:crf}
Q(y_1,...,y_n) = b[y_1] + \sum_{t=1}^{m}s_t[y_t] + \sum_{t=1}^{m-1}T[y_t,y_{t+1}] + \epsilon[y_m]
\end{equation}
where T is a transition matrix $\in$ $R^{10X10}$ and $b,\epsilon \in R^{10}$ are vectors of scores that describe the cost of beginning or ending with a label.

The probability of a sequence of labels is calculated by applying a softmax layer to obtain a probability of a sequence of labels:
\begin{equation}
\label{eq:softmax}
P(y_1,...,y_n) = \frac{e^{Q(y_1,...,y_n)}}{\sum_{y_1,...,y_n}e^{Q(y_1,...,y_n)}}
\end{equation}
Cross-entropy loss, $-log(P(\hat{y}))$, is used as the objective function where $\hat{y}$ is the correct sequence of labels and the probability $P(\hat{y})$ is calculated according to the CRF.

\subsubsection{Running the Model}
We evaluated our model on the 515 annotated history of present illness notes, which were split in a 70\% train set, 15\% development set, and a 15\% test set. The model is trained using the Adam algorithm for gradient-based optimization \cite{kingma2014adam} with an initial learning rate = 0.001 and decay = 0.9. A dropout rate of 0.5 was applied for regularization, and each batch size = 20. The model ran for 20 epochs and was halted early if there was no improvement after 3 epochs.

We evaluated the impact of character embeddings, the choice of pretrained w2v embeddings, and the addition of learned word embeddings on model performance on the dev set. We report performance of the best performing model on the test set. 

\section{Results and Discussion}
\begin{figure}[h!]
\includegraphics[scale=0.35]{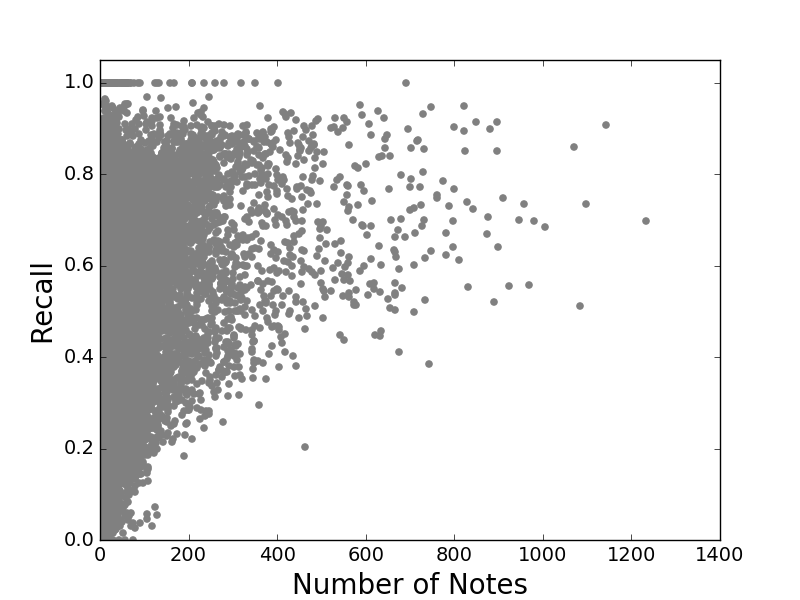}
\includegraphics[scale=0.35]{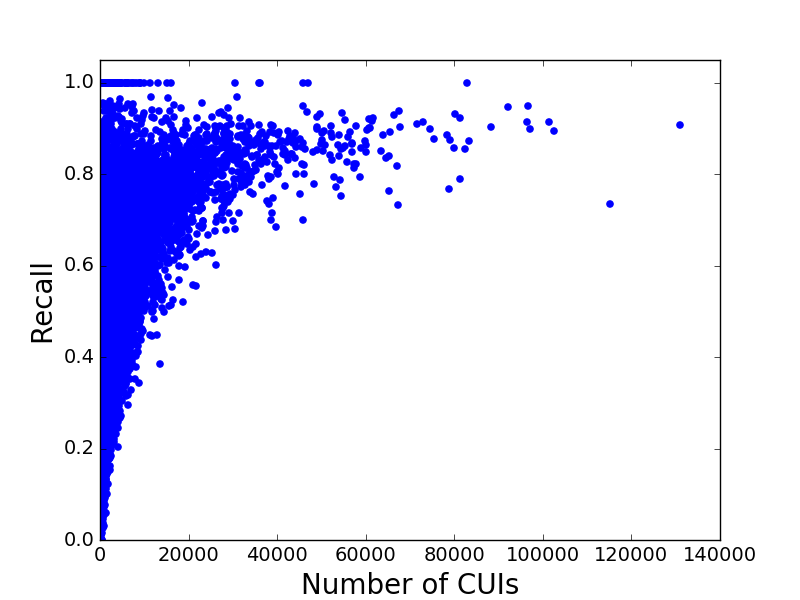}
\includegraphics[scale=0.35]{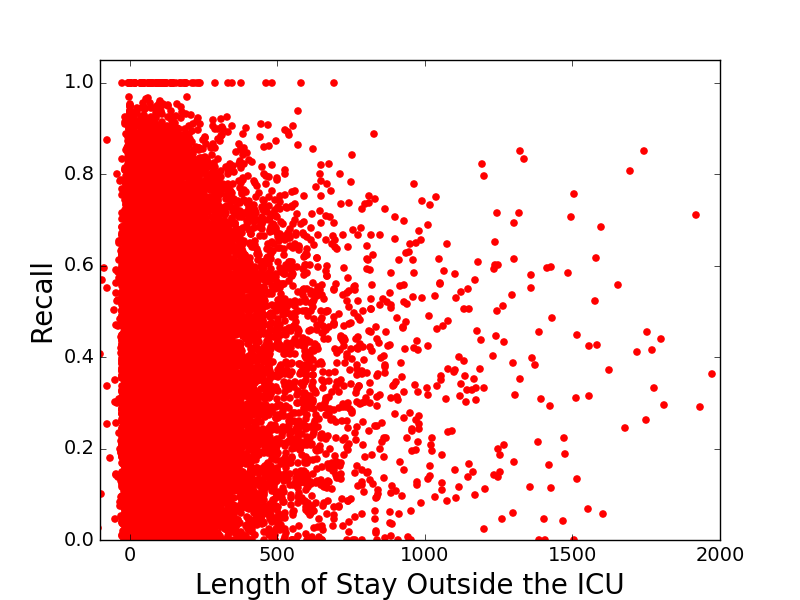}
\centering
\caption{Relationship between the number of non-discharge notes (top), number of non-discharge CUIs (middle), or the number of hours the patient spent outside the ICU (bottom) and the recall for each discharge summary.}
\label{fig:other_notes}
\end{figure}

\subsection{Upper Bound on Summarization}
 For each of the 55,177 discharge summary reports in the MIMIC database, we calculated what fraction of the CUIs in the discharge summary could be found in the remaining notes for the patient's current admission ($\textit{by hadm\_id}$) and in their entire longitudinal medical record ($\textit{by subject\_id}$). Table \ref{tab:recall} shows the CUI recall averaged across all discharge summaries by both subject\_id and hadm\_id. The low recall suggests that clinicians may incorporate information in the discharge note that had not been previously documented in the EHR.  Figure \ref{fig:other_notes} plots the relationship between the number of non-discharge notes for each patient and the CUI recall (top) and the number of total CUIs in non-discharge notes and the CUI recall (middle). The number of CUIs is a proxy for the length of the notes, and as expected, the CUI recall tends to be higher in patients with more and longer notes. The bottom panel in Figure \ref{fig:other_notes} demonstrates that recall is not correlated with the patient's length of stay outside the ICU, which indicates that our upper bound calculation is not severely impacted by only having access to the patient's notes from their stay in the ICU. 
 
 Finally, Table \ref{tab:section_recall} shows the recall for the sex, chief complaint, procedure, discharge medication, and HPI discharge summary sections averaged across all the discharge summaries. The procedure section has the highest recall of 0.807, which is understandable because procedures undergone during an inpatient stay are most likely to be documented in an EHR. The recall for each of these five sections is much higher than the overall recall in Table \ref{tab:recall}, suggesting that extractive summarization may be easier for some sections of the discharge note. 
 
Overall, this upper bound analysis suggests that we may not be able to recreate a discharge summary with extractive summarization alone. While CUI comparison allows for comparing medically relevant concepts, cTAKES's CUI labelling process is not perfect, and further work, perhaps through sophisticated regular expressions, is needed to define the limits of extractive summarization.

\begin{table}[h!]
\caption{Discharge CUI recall for each patient's current hospital admission ($\textit{by hadm\_id}$) and their entire EHR ($\textit{by subject\_id}$), averaged across all discharge summaries}
\label{tab:recall}
\vspace{5pt}
\noindent\begin{tabular}{|p{4.6cm}|p{3.0cm}|} \hline
\textbf{Comparison} & \textbf{Average Recall}  \\ \hline
By subject\_id & 0.431 \\ \hline  
By hadm\_id & 0.375 \\ \hline 
\end{tabular}
\end{table}

\begin{table}[h!]
\caption{Average Recall for five sections of the discharge summary. Recall for each patient's sex was calculated by examining the structured data for the patient's current admission, and recall for the remaining sections was calculated by comparing CUI overlap between the section and the remaining notes for the current admission.}
\label{tab:section_recall}
\vspace{5pt}
\noindent\begin{tabular}{|p{4.6cm}|p{3.0cm}|} \hline
\textbf{Section} & \textbf{Average Recall}  \\ \hline
Sex & 0.675 \\ \hline 
Chief Complaint & 0.720 \\ \hline 
Procedure & 0.807 \\ \hline 
Discharge Medication & 0.580 \\ \hline 
HPI & 0.665 \\ \hline 
\end{tabular}
\end{table}

\subsection{Labeling History of Present Illness Notes}
Table \ref{tab:lstm_f1} compares dev set performance of the model using various pretrained word embeddings, with and without character embeddings, and with pretrained versus learned word embeddings. The first row in each section is the performance of the model architecture described in the methods section for comparison. Models using word embeddings trained on the discharge summaries performed better than word embeddings trained on all MIMIC notes, likely because the discharge summary word embeddings better captured word use in discharge summaries alone. Interestingly, the continuous bag of words embeddings outperformed skip gram embeddings, which is surprising because the skip gram architecture typically works better for infrequent words \cite{word2vecCode}.  As expected, inclusion of character embeddings increases performance by approximately 3\%. The model with word embeddings learned in the model achieves the highest performance on the dev set (0.886), which may be because the pretrained worm embeddings were trained on a previous version of MIMIC. As a result, some words in the discharge summaries, such as mi-spelled words or rarer diseases and medications, did not have associated word embeddings. Performing a simple spell correction on out of vocab words may improve performance with pretrained word embeddings. 

\begin{figure}[h!]
\includegraphics[scale=0.38]{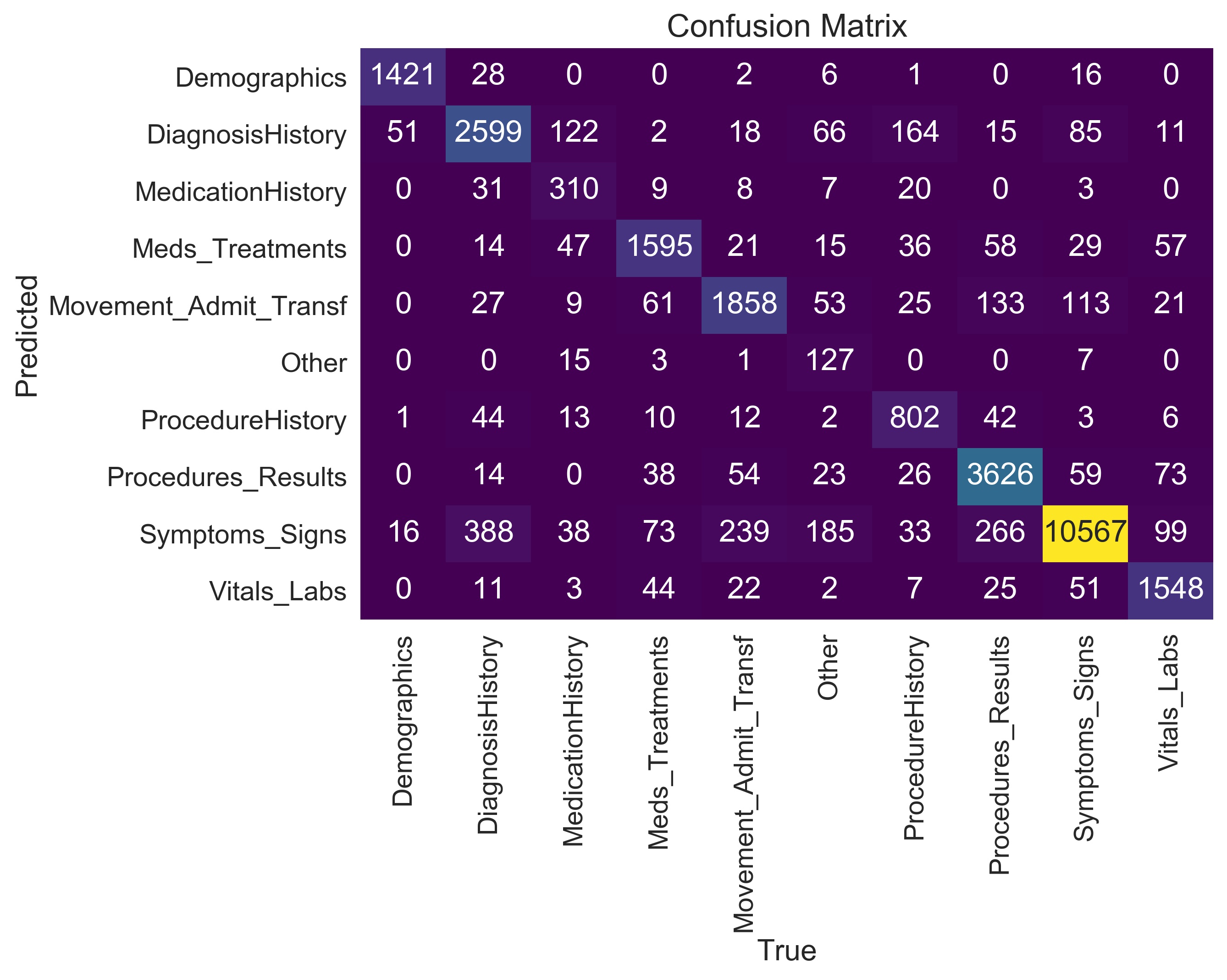}
\centering
\caption{Confusion Matrix for the Learned Word Embedding model on the test set. True labels are on the x-axis, and predicted labels are on the y-axis.}
\label{fig:conf_matrix}
\end{figure}

We evaluated the best performing model on the test set. The Learned Word Embeddings model achieved an accuracy of 0.88 and an F1-Score of 0.876 on the test set. Table \ref{tab:lstm_prec_recall} shows the precision, recall, F1 score, and support for each of the ten labels, and Figure \ref{fig:conf_matrix} shows the confusion matrix illustrating which labels were frequently misclassified. The demographics and patient movement labels achieved the highest F1 scores (0.96 and 0.93 respectively) while the vitals/labs and medication history labels had the lowest F1 scores (0.40 and 0.66 respectively). The demographics section consistently occurs at the beginning of the HPI note, and the patient movement section uses a limited vocab (transferred, admitted, etc.), which may explain their high F1 scores. On the other hand, the vitals/labs and medication history sections had the lowest support, which may explain why they were more challenging to label. 

Words that belonged to the diagnosis history, patient movement, and procedure/results sections were frequently labeled as symptoms/signs (Figure \ref{fig:conf_matrix}). Diagnosis history sections may be labeled frequently as symptoms/signs because symptoms/diseases can either be described as part of the patient's diagnosis history or current symptoms depending on when the symptom/disease occurred. However, many of the misclassification errors may be due to inconsistency in manual labelling among annotators. For example, sentences describing both patient movement and patient symptoms (e.g. "the patient was transferred to the hospital for his hypertension") were labeled entirely as 'patient movement' by some annotators while other annotators labeled the different clauses of the sentence separately as 'patient movement' and 'symptoms/signs.' Further standardization among annotators is needed to avoid these misclassifications. Future work is needed to obtain additional manual annotations where each HPI note is annotated by multiple annotators. This will allow for calculation of Cohen's kappa, which measures inter-annotator agreement, and comparison of clinician and non-clinician annotator reliability. 

Future work is also needed to better understand commonly mislabeled categories and explore alternative model architectures. Here we perform word level label prediction, which can result in phrases that contain multiple labels. For example, the phrase "history of neck pain" can be labeled with both 'diagnosis history' and 'symptoms/signs' labels. Post-processing is needed to create a final label prediction for each phrase. While phrase level prediction may resolve these challenges, it is difficult to segment the HPI note into phrases for prediction, as a single phrase may truly contain multiple labels. Segmentation of sentences by punctuation, conjunctions, and prepositions may yield the best phrase chunker for discharge summary text. 

Finally, supplementing the word embeddings in our LSTM model with CUIs may further improve performance. While word embeddings do well in learning the contextual context of words, CUIs allow for more explicit incorporation of medical domain expertise. By concatenating the CUI for each word with its hybrid token embedding, we may be able to leverage both data driven and ontology driven approaches.  

\begin{table}[h!]
\caption{F1-Scores on the development set under various model architectures. The first model in each section is the same model described in the methods section. Top performing models in each section are in bold.}
\vspace{5pt}
\label{tab:lstm_f1}
\noindent\begin{tabular}{|p{6.3cm}|p{1.4cm}|} \hline
\textbf{Model} & \textbf{F1}  \\ \hline
w2v CBOW; Discharge Summaries & \textbf{0.873} \\ \hline
w2v Skip Gram; Discharge Summaries & 0.831  \\ \hline
w2v CBOW; All MIMIC Notes & 0.862  \\ \hline
w2v Skip Gram; All MIMIC Notes & 0.809  \\ \hline \hline
With Character Embeddings & \textbf{0.873}\\ \hline
Without Character Embeddings & 0.847\\ \hline \hline
Pretrained Word Embeddings & 0.873\\ \hline
Learned Word Embeddings & \textbf{0.886}\\ \hline 
\end{tabular}
\end{table}

\begin{table}[h!]
\caption{ Precision (P), Recall (R), F1 Score, and Number of supporting examples for each of the 10 label categories for the Learned Word Embedding model on the test set.}
\vspace{5pt}
\label{tab:lstm_prec_recall}
\noindent\begin{tabular}{|p{3.2cm}|p{0.7cm}|p{0.7cm}|p{0.7cm}|p{1.1cm}|} \hline
Label & P & R & F1 & Support \\ \hline
Demographics & 0.96 & 0.95 & 0.96 & 1489 \\
DiagnosisHistory &0.83 &0.82 & 0.83 & 3156 \\
MedicationHistory & 0.80 & 0.56 & 0.66 & 557 \\
ProcedureHistory & 0.85 & 0.87 & 0.86 & 1835\\
Symptoms/Signs & 0.81 & 0.83 & 0.82 & 2235 \\
Vitals/Labs & 0.83 & 0.26 & 0.40 & 486\\
Procedures/Results & 0.86 & 0.72 & 0.78 & 1114 \\
Meds/Treatments & 0.93 & 0.87 & 0.90 & 4165\\
PatientMovement & 0.89 & 0.97 & 0.93 & 10933\\
Other & 0.90 & 0.85 & 0.88 & 1815\\ \hline
\textbf{Average/Total} & \textbf{0.88} & \textbf{0.88} & \textbf{0.88} & \textbf{27785}\\ \hline
\end{tabular}
\end{table}

\section{Conclusion}
In this paper we developed a CUI-based upper bound on extractive summarization of discharge summaries and presented a NN architecture that jointly classifies words in history of present illness notes. We demonstrate that our model can achieve excellent performance on a small dataset with known heterogeneity among annotators. This model can be applied to the 55,000 discharge summaries in MIMIC to create a dataset for evaluation of extractive summarization methods.

\section*{Acknowledgments} 
We would like to thank our annotators, Andrew Goldberg, Laurie Alsentzer, Elaine Goldberg, Andy Alsentzer, Grace Lo, and Josh Donis. We would also like to acknowledge Pete Szolovits for his guidance and for providing the pretrained word embeddings and Tristan Naumann for providing the MIMIC CUIs. 

\bibliography{paper}
\bibliographystyle{icml2013}

\end{document}